\documentclass[prd,showpacs,letterpaper,twocolumn]{revtex4}%
\usepackage{graphicx}
\usepackage{bm}
\usepackage{epsf}
\usepackage{rotating}
\usepackage{epsfig,graphics,rotate,color}
\usepackage{wrapfig}
\usepackage{amssymb}
\usepackage{amsmath}
\usepackage{amsfonts}
\usepackage{array,hhline,dcolumn}%
\providecommand{\U}[1]{\protect\rule{.1in}{.1in}}
\begin{document}
\title{Problems With the MINOS/MINOS+ Sterile
  Neutrino $\nu_\mu$ Disappearance Result}
\author{W. C. Louis$^{1}$}
\affiliation{$^{1}$ Los Alamos National Laboratory; Los Alamos, NM 87545}

\date{\today}

\begin{abstract}

The MINOS/MINOS+ experiment has recently reported stringent limits on $\nu_\mu$ 
disappearance that appear to rule out the 3+1 sterile neutrino model. However,
in this paper we wish to point out problems associated 
with the MINOS/MINOS+ analysis. In particular, we find that MINOS/MINOS+ has either 
underestimated their systematic errors and/or has obtained evidence for physics
beyond the 3-neutrino paradigm. Either case would invalidate the limits on 
$\nu_\mu$ disappearance.

\end{abstract}

\maketitle

\section{Introduction}

The MINOS/MINOS+ collaboration has recently presented stringent limits
on short-baseline $\nu_\mu$ disappearance \cite{minos} in the $\Delta m^2$ region
from $0.0001$ eV$^2$ to 1000 eV$^2$. At moderate values of $\Delta m^2$
around 1 eV$^2$, the experimental procedure makes use of the relative 
rate of neutral current (NC) and charged current (CC) events in the far and near 
detectors. At high values of $\Delta m^2$
around 1000 eV$^2$, the near and far detectors will see identical 
oscillation effects that will vanish in the ratio. In that case, the 
experiment must rely upon comparing neutrino data to an absolute 
prediction, which we will call ``dead reckoning'' in this article.     

However, the MINOS/MINOS+ analysis has two significant problems that make the
limits dramatically too good.
First, the systematic uncertainties used in the analysis appear to
be much too low. Second, there appears to be an unknown systematic bias that
results in relatively more NC events in the far detector than in
the near detector.

\section{Systematic Uncertainty Problem}

\begin{figure}
\begin{center}
\includegraphics[width=0.55\textwidth]{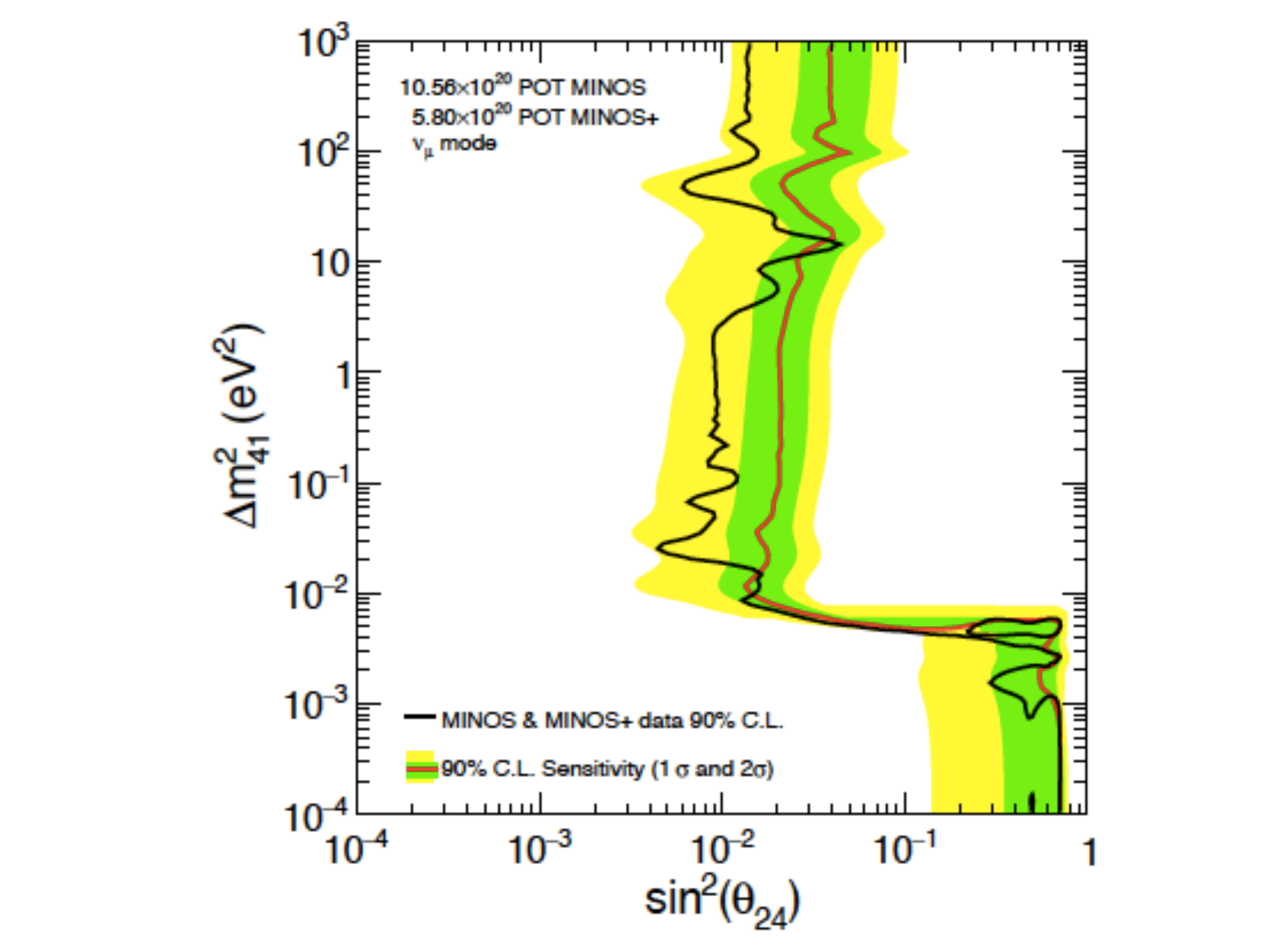}
  \caption{\small The MINOS/MINOS+ exclusion limit from reference \cite{minos}.}
  \label{pluslimit}
  \end{center}
\end{figure}

The MINOS/MINOS+ limit plot is shown in Fig.~\ref{pluslimit}. 
At 1000 eV$^2$, the limit (sensitivity) for $\sin^2\theta_{24}$
is $\sim 1.2\%$ ($\sim 4\%$). This corresponds to a fractional
error for the absolute, dead-reckoning prediction of the event rate to be
$\sim 2\%$ ($\sim 6\%$), which is too low to be credible. The total systematic
error needs to include the systematic errors on the neutrino flux, the neutrino 
cross section, the detector efficiency, the DAQ efficiency, and many other 
contributions, and it seems impossible that the cumulative value for all
of these systematic errors is 2\% (6\%). Based on the experience of other neutrino 
experiments around the world, one would expect a total systematic error $>15\%$.

\section{Systematic Bias Problem}

\begin{figure}
\begin{center}
\includegraphics[width=0.55\textwidth]{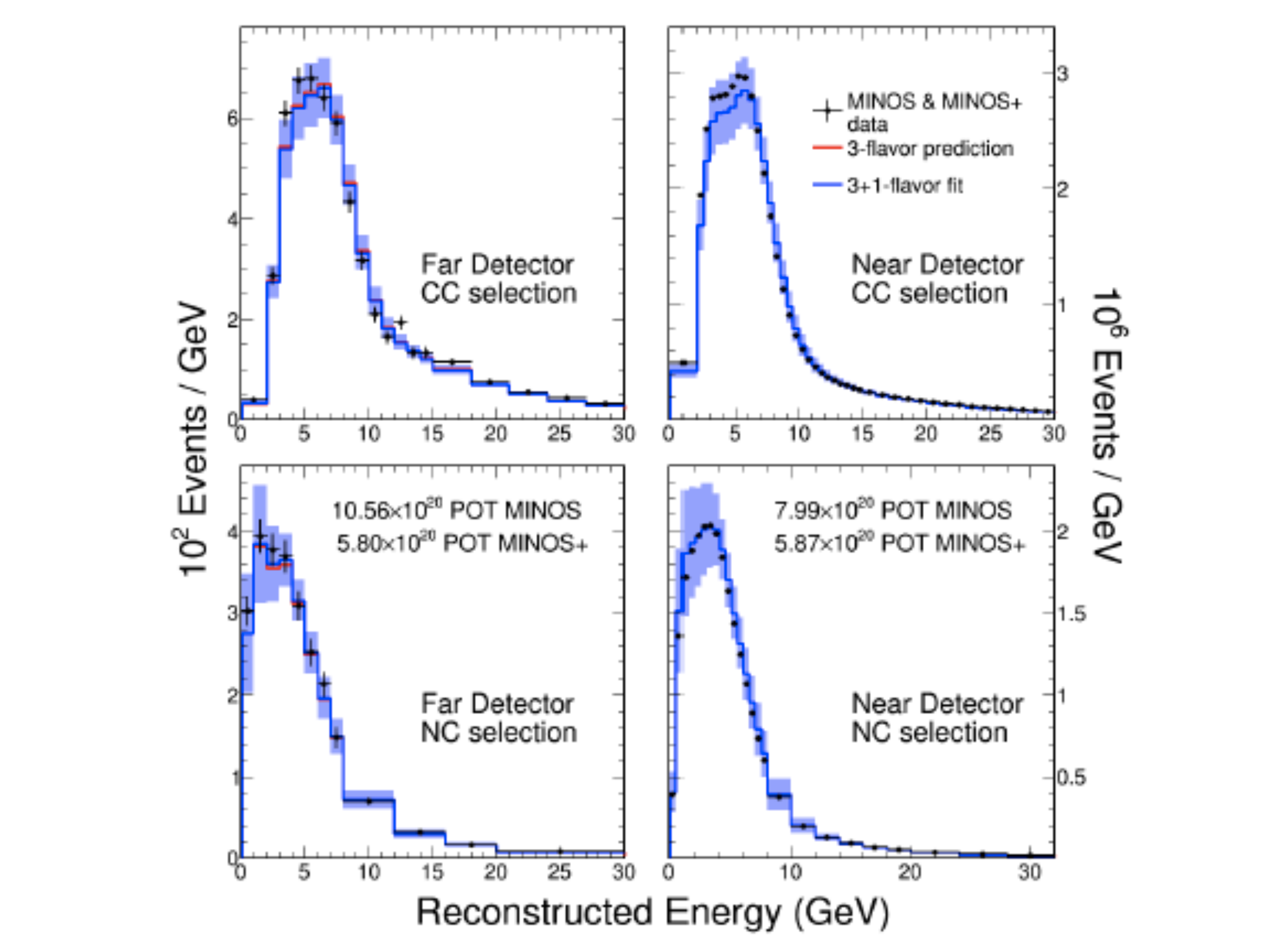}
  \caption{\small The summed MINOS/MINOS+ reconstructed energy spectra from reference \cite{minos}.}
  \label{datamc}
  \end{center}
\end{figure}

The MINOS/MINOS+ data-to-predicton plots in Fig.~\ref{datamc} show the ratio of
data to the no-oscillation prediction for, from left to right, CC far
detector, NC far detector, CC near detector and NC near detector.
As can be seen in the figure, the NC data events appear to be above (below) the 
Monte Carlo 
prediction in the far (near) detector, which implies that there are relatively more
NC events in the far detector than the near detector. To quantify this effect, we
calculate the ratio of ratios, $R$, to be the ratio of NC
events observed to expected in the far detector compared to the
near detector. A fit to the data, using only statistical uncertainties, yields
$R = 1.062\pm 0.019$, which corresponds to a 3.3 $\sigma$ statistical effect.   
The $\chi^2 = 21.1/12$ DF for $R=1$ and $\chi^2 = 10.1/11$ DF for $R=1.062$, giving 
$\Delta \chi^2 = 11.0$ or a probability of $9\times 10^{-4}$ for the expected value of 
$R=1$. This clearly shows
that either there is a large systematic error and/or there is evidence for
neutrino appearance from physics beyond the 3-neutrino paradigm.
Note that this cannot be explained by 3-active neutrino oscillations.
Note also that the MINOS/MINOS+ analysis assumes that the probability of 
$\nu_\mu \rightarrow \nu_e$ oscillations is zero.
However, in 3+N models with three active neutrinos and N sterile neutrinos, there
will, in general, be both $\nu_\mu$ disappearance and $\nu_e$ appearance.

\section{Conclusions}

In summary, we note the following two problems with
the MINOS/MINOS+ analysis.  

\begin{itemize}

\item The limit (sensitivity) at high  $\Delta m^2$ indicates an
$\sim 2\%$ ($\sim 6\%$) uncertainty on the absolute, dead-reckoning determination
of the event rates in the near detector,  which is
not credible given that one would expect an uncertainty $>15\%$ based on other
neutrino experiments.

\item The ratio of NC
events observed to expected in the far detector to the
near detector is $1.062 \pm 0.019$, assuming statistical uncertainties.
This is $\sim 3.3 \sigma$ from unity. 

\end{itemize}

MINOS/MINOS+ has either
underestimated their systematic errors and/or has obtained evidence for physics
beyond the 3-neutrino paradigm. Either case would invalidate the limits on
$\nu_\mu$ disappearance.


\begin{thebibliography}{99}                                                                   

\bibitem{minos} P. Adamson et al., arXiv:1710.06488.

\end{thebibliography}
\end{document}